\documentclass[fleqn,usenatbib,pdfpagelabels=false,pdfstartview=XYZ, bookmarks=true, colorlinks=true, linkcolor=blue, urlcolor=blue, citecolor=blue, pdftex, bookmarks=true, linktocpage=true, hyperindex=true]{mnras}

\usepackage{newtxtext,newtxmath}
\usepackage[T1]{fontenc}

\DeclareRobustCommand{\VAN}[3]{#2}
\let\VANthebibliography\thebibliography
\def\thebibliography{\DeclareRobustCommand{\VAN}[3]{##3}\VANthebibliography}

\usepackage{graphicx}	% Including figure files
\usepackage{amsmath}	% Advanced maths commands
\usepackage{academicons}
\usepackage{orcidlink}

\newcommand{\orcid}[1]{\href{https://orcid.org/#1}{\textcolor[HTML]{A6CE39}{\aiOrcid}}}

\title[The 3D Geometry of IC\,59, IC\,63, and Gamma Cas]{The 3D Geometry of Reflection Nebulae IC\,59 and IC\,63 with their illuminating Star Gamma Cas}

\author[Jacob M. Eiermann]{
Jacob M. Eiermann$^{1}\thanks{Contact E-mail: Eiermannj7@gmail.com} \orcidlink{0000-0001-9728-4262}$
Miranda Caputo$^{2} \orcidlink{0000-0002-2957-3924}$
Thomas S.-Y. Lai$^{3} \orcidlink{0000-0001-8490-6632}$
Adolf N. Witt$^{2} \orcidlink{0000-0003-0760-4483}$
\\
$^{1}$Department of Physics $\&$ Astronomy, University of Georgia, Sanford Drive, Athens, GA 30602, USA
\\
$^{2}$Ritter Astrophysical Research Center, University of Toledo, Toledo, OH 43606, USA
\\
$^{3}$IPAC, California Institute of Technology, 1200 E. California Blvd., Pasadena, CA 91125, USA}
\date{Accepted 01/04/2024. Received YYY; in original form ZZZ}

\pubyear{2023}

\begin{document}
\label{firstpage}
\pagerange{\pageref{firstpage}--\pageref{lastpage}}
\maketitle
%\footnote{Eiermannj7@gmail.com}

%% Note that the \and command from previous versions of AASTeX is now
%% depreciated in this version as it is no longer necessary. AASTeX 
%% automatically takes care of all commas and "and"s between authors names.

%% AASTeX 6.31 has the new \collaboration and \nocollaboration commands to
%% provide the collaboration status of a group of authors. These commands 
%% can be used either before or after the list of corresponding authors. The
%% argument for \collaboration is the collaboration identifier. Authors are
%% encouraged to surround collaboration identifiers with ()s. The 
%% \nocollaboration command takes no argument and exists to indicate that
%% the nearby authors are not part of surrounding collaborations.

%% Mark off the abstract in the ``abstract'' environment. 
\begin{abstract}

The early-type star $\gamma$\,Cas illuminates the reflection nebulae IC\,59 and IC\,63, creating two photo-dissociation regions (PDRs). Uncertainties about the distances to the nebulae and the resulting uncertainty about the density of the radiation fields incident on their surfaces have hampered the study of these PDRs during the past three decades. We employed far-UV -- optical nebula -- star colour differences of dust-scattered light to infer the locations of the nebulae relative to the plane of the sky containing $\gamma$\,Cas, finding IC\,63 to be positioned behind the star and IC\,59 in front of the star. To obtain the linear distances of the nebulae relative to $\gamma$\,Cas, we fit far-infrared archival \textit{Herschel} flux data for IC\, 59 and IC\, 63 with modified blackbody (MBB) curves and relate the resulting dust temperatures with the luminosity of $\gamma$\,Cas, yielding approximate distances of 4.15 pc for IC\,59 and 2.3 pc for IC\,63. With these distances, using updated far-UV flux data in the 6 eV – 13.6 eV range for $\gamma$\,Cas with two recent determinations of the interstellar extinction for $\gamma$\,Cas, we estimate that the far-UV radiation density at the surface of IC\,63 takes on values of $G_0$ = 58  or $G_0$ = 38 with respective values for E(B-V) for $\gamma$\,Cas of 0.08 and 0.04 mag. This is a substantial reduction from the range 150 $\le$ $G_0$ $\le$ 650 used for IC\,63 during the past three decades. The corresponding, even lower new values for IC\,59 are $G_0$ = 18 and $G_0$ = 12.

\end{abstract}


\section{Introduction} \label{sec:intro}

\begin{figure}
    %\centering
    \includegraphics[width=0.40\textwidth]{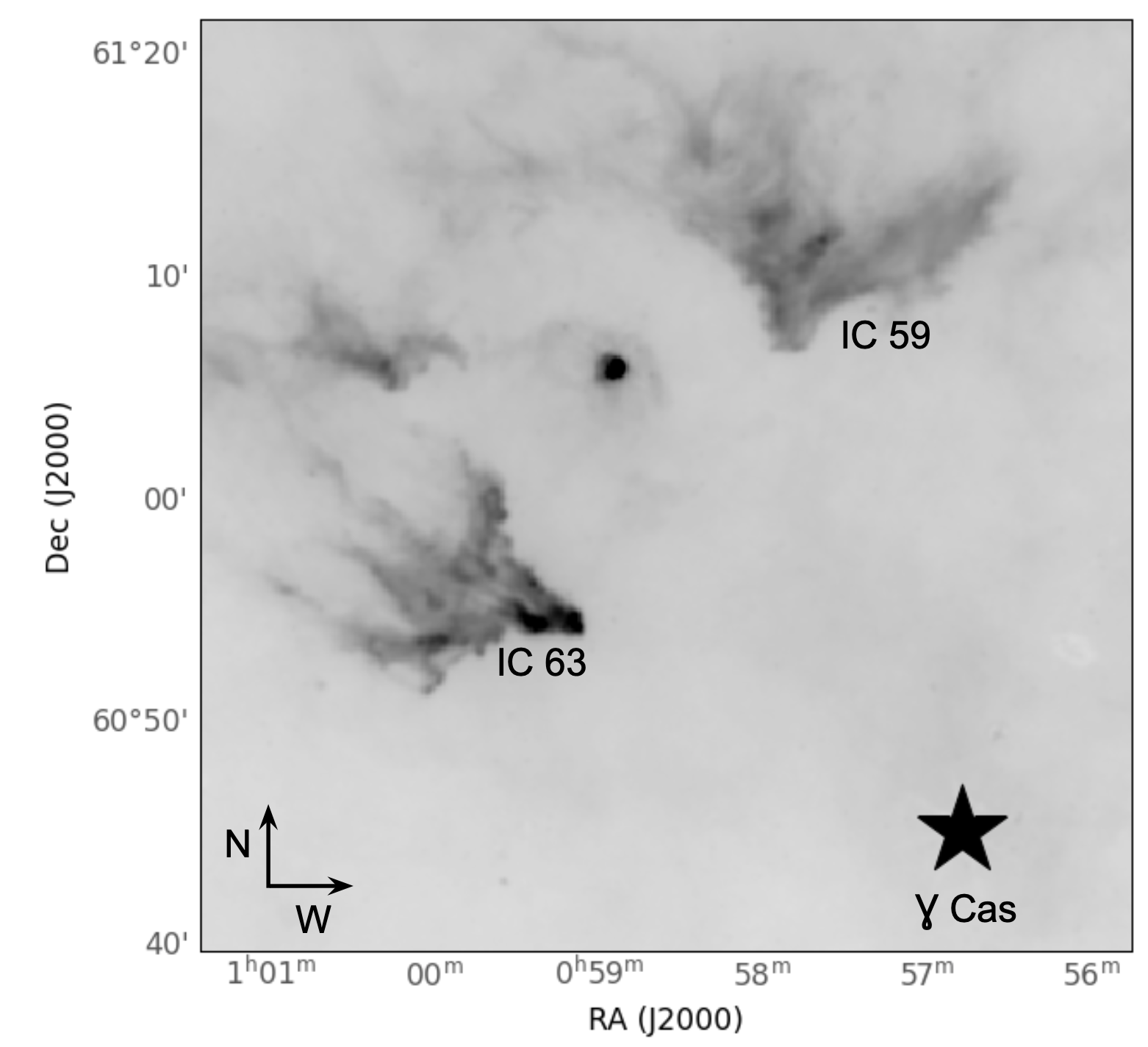}
    \caption{WISE Band 3 (12$\mu$m) cutout of region containing $\gamma$\,Cas with IC\,59 and IC\,63.}
    \label{FIG:1}
\end{figure}

IC\,59 and IC\,63 are two reflection/emission nebulae that are externally illuminated by the bright B0.5IVpe star $\gamma$\,Cas (Figure \ref{FIG:1}). Sculpted by the UV radiation from $\gamma$\,Cas, the nebulae appear as cometary neutral globules in a larger cavity of ionized gas surrounding the star \citep{2021ApJ...922..183C}, the H\,II region Sh2-185 \citep{1997MNRAS.287..455B}. Their relative proximity to Earth (d = (168 $\pm$ 4) pc; \citet{2007A&A...474..653V}) makes them ideal laboratories for studying the interaction of intense stellar UV radiation with molecules and dust, i.e., where models for photo-dissociation regions (PDRs) can be tested with the highest spatial resolution under conditions, where the PDR stratification can be clearly seen.

IC\,63, the brighter of the two, has been the object of more extensive observation and study. It was this nebula where UV fluorescence emission of molecular hydrogen was first observed outside the solar system \citep{1989ApJ...336L..21W,1989ApJ...347..863S,1998ApJ...500L..61H,2005ApJ...628..750F}. 
This discovery was followed in rapid succession by numerous other investigations at longer wavelengths directed at IC\,63 as well as IC\,59, with a focus on the physics and chemistry probed by a diverse array of molecules with their emission mechanisms
\citep{1994A&A...282..605J, 1995A&A...302..223J, 1996A&A...309..899J, 1997MNRAS.287..455B, 2004A&A...414..531H, 2005AJ....129..954K, 2005ApJ...628..750F, 2009MNRAS.400..622T, 2010ApJ...725..159F, 2010ApJ...717..658M, 2013ApJ...775...84A, 2017MNRAS.469.4933L, 2018A&A...619A.170A, 2017MNRAS.465..559S, 2021ApJ...923..107S, 2021ApJ...919...27K, 2020ApJ...888...22V, 2022A&A...666A..49S, 2023ApJ...950..140C, 2023AJ....165..243B}.

A persistent problem facing most of these investigations arose from the fact that the spatial distance from $\gamma$\,Cas to the two nebulae was highly uncertain. The analysis of the data in PDR studies requires information about the density of the stellar UV radiation field incident on the nebulae. Lacking information about the positions of the nebulae along their respective lines of sight, most studies assumed that the nebulae were situated in the plane of the sky that contains $\gamma$\,Cas, i.e., that the angular offsets from the star at the star’s distance determined the physical distances. With the currently accepted distance to $\gamma$\,Cas of (168$\pm$4) pc, this offset distance amounts to about 1.0 pc in the case of IC\,63 and 1.2 pc for IC\,59. This resulted in the frequently used assumption that the UV radiation flux in the energy range 6 eV - 13.6 eV incident on the face of IC\,63 amounts to about $G_0$ = 650 Draine units \citep{1994A&A...282..605J}, where one Draine unit is 2.7 x $10^{-3}$ erg s$^{-1}$ cm$^{-2}$ \citep{1978ApJS...36..595D}, or about 1100 \citet{1968BAN....19..421H} units.

\citet{1995A&A...302..223J} was first in considering possible geometries where IC\,63 was located behind $\gamma$\,Cas, with the line toward IC\,63 forming an angle of 30$^{\circ}$ with the plane of the sky at the location of $\gamma$\,Cas. This would increase the distance to IC\,63 to about 1.15 pc. This suggestion was pursued further by \citet{2013ApJ...775...84A}, who found that the angular offset between linear streaming features seen in maps of the 1-0 S(1) emission of H$_{2}$ and the direction of the incident radiation suggested instead an angle of 68$^{\circ}$ with the plane of the sky, placing IC\,63 further behind. The linear distance from $\gamma$\,Cas to IC\,63 in this instance is about 2.7 pc. Analyzing available data from the \textit{Spitzer} and \textit{Herschel} programs, \citet{2018A&A...619A.170A} concluded that the tips of IC\,63 and IC\,59 were, respectively, about three and five times farther away from $\gamma$\,Cas than their projected angular offset distances, but with no indication as to the location of the nebulae relative to the plane of the sky containing $\gamma$\,Cas. They recommended $G_0$ = 150 for IC\,63 and $G_0$ = 25 for IC\,59. Most recently, \citet{2023ApJ...950..140C}, from a study of [C II] kinematics in both nebulae, concluded that IC\,63 is located behind the plane of the sky containing $\gamma$\,Cas, at about 2 pc from $\gamma$\,Cas, whereas IC\,59 is placed in front of this plane at about 4.5 pc from $\gamma$\,Cas, being irradiated at an angle of about 70$^{\circ}$.

In this work, we used the observed surface brightness of dust-scattered light at optical and UV wavelengths and corresponding measurements of the flux from $\gamma$\,Cas at earth to determine the nebula - star colour differences for IC\,63 and IC\,59. 
Such colour differences are sensitive to changes in the effective scattering angle, because the scattering phase function of interstellar dust is strongly forward-throwing at UV-optical wavelengths, with the phase function asymmetry increasing toward shorter wavelengths \citep{1985ApJ...294..216W, 2003ApJ...598.1017D, 2022A&A...659A.149B}. 
Previously observed colour differences as a function of offset distance in the Merope nebula have demonstrated the magnitude and direction of this effect \citep{1977A&A....54..841A, 1977PASP...89..750W}. 
Thus, colour-difference measurements in IC\,59 and IC\,63 are useful in determining the locations of the nebulae along their lines of sight relative to $\gamma$\,Cas. 
Additionally, we have used archival far-IR \textit{Herschel} data covering the 70 $\mu$m - 500 $\mu$m wavelength range to estimate the temperatures of large grains in equilibrium with the radiation field of $\gamma$\,Cas. 
With information about the star’s bolometric luminosity and assuming that large grains (approximate size range 100 nm -- 1000 nm) are in equilibrium with the radiation field of $\gamma$\,Cas, these temperatures yield estimates of the radial distance of the nebulae from $\gamma$\,Cas, independent of direction. 
Further constraints on the ratio of these two distances come from measurements of the maximum intensities of extended red emission (ERE, \citet{2017MNRAS.469.4933L}) in both nebulae. 
The ERE appears as an isotropic emission in the 5400 Å - 8500 Å wavelength range in many interstellar environments containing interstellar dust and far-UV photons \citep{2020Ap&SS.365...58W}. 
The ERE intensity has been shown to correlate with the density of the prevailing far-UV (10.5 eV - 13.6 eV) radiation field \citep{2002ApJ...565..304S, 2006ApJ...636..303W} and can therefore be employed to estimate the density of such fields. 
Finally, we used the estimated linear distances from $\gamma$\,Cas to the two nebulae, together with far-UV \textit{IUE} and \textit{FUSE} spectra and existing constraints on the interstellar reddening of $\gamma$\,Cas, to calculate the most likely UV radiation densities at the faces of IC\,59 and IC\,63.

We record the sources of our data in Section 2, followed by the analysis of the different data sets in Section 3. In Section 4 we discuss the limitations and uncertainties of our results arising from various assumptions made in the analysis, followed by a set of conclusions in Section 5.

\section{Data} \label{sec:data}
\subsection{Distance and Luminosity of $\gamma$\,Cas}

We adopted the distance of (168 $\pm$ 4) pc for $\gamma$\,Cas \citep{2007A&A...474..653V}, which is a reduction from the original Hipparcos distance of (188 $\pm$ 20) pc \citep{1997A&A...323L..49P}, resulting from the corrections for systematic errors present in the original catalog. Most earlier investigations of the $\gamma$\,Cas/IC\,63/IC\,59 system have assumed distances in the 230 pc \citep{1994A&A...282..605J} to 190 pc \citep{2023ApJ...950..140C} range.

Setting aside the evidence for its long-term and cyclical variability (\citet{2012ApJ...760...10H}, and references therein), \citet{2007ApJ...668..481S} adopted the value L$_{\star}$ = 34,000 L$_{\odot}$ for the present luminosity of $\gamma$\,Cas, with an assumed distance d = 188 pc. For our adopted distance of d = 168 pc, this reduces to L$_{\star}$ = 27,150 L$_{\odot}$, which we will use in our subsequent calculations.

\subsection{Interstellar Extinction}
$\gamma$\,Cas is the well-studied prototype of classical Be stars. The colour excess E(B-V) of Be stars is in part the result of free-free emission of their circumstellar disks affecting their V-magnitudes and is, therefore, not a suitable basis for estimating the interstellar extinction for such stars \citep{1983A&A...120..223S}. Nonetheless, interstellar absorption features such as the 2175 Å absorption band and diffuse interstellar bands (DIBs) that closely correlate with interstellar reddening have led to recent estimates of the interstellar reddening for $\gamma$\,Cas. These values are E(B-V) = (0.08 $\pm$ 0.007) mag, based on the equivalent widths of the DIBs at 5780 Å and 5797 Å \citep{2017BlgAJ..27...10N}, and E(B-V) = $0.04^{+0.03}_{-0.02}$ mag \citep{2017A&A...601A..74K}, arrived at by canceling the 2175 Å band absorption in the ultraviolet SED of $\gamma$\,Cas with an average galactic extinction curve of \citet{1989ApJ...345..245C}, hereafter referred to as CCM, with a ratio of total-to-selective extinction $R_V$ = 3.1. We, therefore, adopt these reddening values as limiting values, and we further assume that the extension of the extinction curve into the UV follows a CCM curve with $R_V$ = 3.1. Furthermore, we will assume that the light originating from IC\,59, IC\,63, and $\gamma$\,Cas is subjected to the same amount of foreground interstellar extinction.

\subsection{Surface Brightness Data for IC\,59 and IC\,63}
\label{sec:2_3}
We acquired low-resolution, large-aperture (10" x 20") \textit{IUE} spectra of IC\,59 and IC\,63 from the \textit{Mikulski Archive for Space Telescopes (MAST)}. There are two spectra associated with IC\,59 that were taken at different positions (RA: 00:57:42.431, DEC: +61:04:59.89 and RA: 00:57:14.476, DEC: +61:07:28.34) within that nebula; we selected the second, brighter of the two for its more favorable S/N ratio. For both IC \,59 and IC\,63, we integrated the nebular scattered-light fluxes in two bands, 2600 Å - 2800 Å and 1750 Å - 1900 Å. We chose these bands for their relatively high S/N ratio in the first case, and in the second instance, to avoid contamination by the far-UV fluorescence of H$_{2}$ that dominates the spectrum of IC\,63 at wavelengths shorter than 1650 Å \citep{1989ApJ...336L..21W}. We note that fluxes observed from extended sources, such as IC 59 and IC 63, depend on the limited solid angles of the apertures of the IUE spectrometers, projected onto the plane of the sky. Therefore, the nebular surface brightness becomes a more meaningful physical measure of the nebular brightness. To convert the fluxes, expressed in units of (erg cm$^{-2}$ s$^{-1}$ Å$^{-1}$) into surface brightness units of (erg cm$^{-2}$ s$^{-1}$ Å$^{-1}$ sr$^{-1}$), we adopted the following solid angles subtended by the large \textit{IUE} apertures: SWLA = 5.03 x $10^{-9}$ sr for the short-wavelength spectrometer, and LWLA = 5.19 x $10^{-9}$ sr for the long-wavelength spectrometer \citep{1980A&A....85....1B} for the normalization to unit steradian. The resulting nebular surface brightness values in the two UV bands are listed in Table \ref{table:1}.

\begin{table*}
\centering
\begin{tabular}{ |c|c|c|c|c|c|  }
 \hline
 \multicolumn{2}{|c|}{} &\multicolumn{2}{|c|}{Position} &\multicolumn{1}{|c|}{Wavelength}&\multicolumn{1}{|c|}{Surface Brightness}  \\
 \hline
 Object Name&\textit{IUE} Spec File ID&R.A.(2000.0)&DEC(2000.0)&[Å]&[erg cm$^{-2}$ s$^{-1}$ Å$^{-1}$ sr$^{-1}$]\\
 \hline
 IC\,63&LWP17444&00:59:01.365&+60:53:17.56&2600 - 2800&(1.13 $\pm$ 0.12)E-06\\
 IC\,63&SWP33989&00:59:01.365&+60:53:17.56&1750 - 1900&(2.11 $\pm$ 0.35)E-06\\
 IC\,59&LWP17252&00:57:14.476&+61:07:28.34&2600 - 2800&(7.42 $\pm$ 1.80)E-07\\
 IC\,59&SWP38087&00:57:14.476&+61:07:28.34&1750 - 1900&(1.81 $\pm$ 0.43)E-06\\
 \hline
\end{tabular}
\caption{Ultraviolet surface brightness measurements from \textit{IUE} spectra}
\label{table:1}
\end{table*}

For corresponding nebular scattered-light surface brightness values in the optical B(4410 Å) and G(5238 Å) bands, at the positions defined by the \textit{IUE} pointing in the two nebulae, we referred to measurements published by \citet{2017MNRAS.469.4933L}. We did not use data from the R (6464 Å) band, given that the presence of ERE in this band significantly contaminates the dust-scattered light in both nebulae. The data of \citet{2017MNRAS.469.4933L} (Table A1 and A2) are flux-calibrated intensities S, divided by the flux observed from $\gamma$\,Cas, F$^{\star}$, in the two bands, listed together with the equatorial coordinates of the corresponding positions in the two nebulae. We interpolated and extrapolated from several of these \citet{2017MNRAS.469.4933L} positions closest to the \textit{IUE} positions to arrive at best estimates for log(S/F$^{\star}$)$_B$ and log (S/F$^{\star}$)$_G$. The results are listed in Table \ref{table:2}, together with corresponding values for the two UV bands (see Sect. 2.4).

\begin{table*}
\centering
\begin{tabular}{ |c|c|c|c|c|c|c|  }
 \hline
 \multicolumn{1}{|c|}{} &\multicolumn{2}{|c|}{Position} &\multicolumn{4}{|c|}{}  \\
 \hline
 Object Name&R.A.(2000.0)&DEC(2000.0)&log(S/$\,F^\star$)$_{1825}$&log(S/$\,F^\star$)$_{2700}$&log(S/$\,F^\star$)$_B$&log(S/$\,F^\star$)$_G$\\
 \hline
 IC\,63&00:59:01.365&+60:53:17.56&2.46 $\pm$ 0.03&2.55 $\pm$ 0.06&2.87 $\pm$ 0.02&2.99 $\pm$ 0.01 \\
 IC\,59&00:57:14.476&+61:07:28.34&2.39 $\pm$ 0.07&2.36 $\pm$ 0.07&2.40 $\pm$ 0.02&2.45 $\pm$ 0.01\\
 \hline
\end{tabular}
\caption{Measured log(S/F$^\star$) in units of [sr$^{-1}$].}
\label{table:2}
\end{table*}

\subsection{Far-UV Flux from $\gamma$\,Cas}
We needed far-UV fluxes for $\gamma$\,Cas for two purposes: One, to compute stellar-flux-normalized nebular surface brightness values log(S/F$^{\star}$), two, to provide a basis from which to calculate the stellar UV fluxes in the energy range 6 eV - 13.6 eV incident on the surfaces of IC\,59 and IC\,63, once their distances from $\gamma$\,Cas are determined.

For the first, we acquired low-resolution, large-aperture (10" x 20") \textit{IUE} $\gamma$\,Cas spectra LWP13607 and SWP33895 from \textit{MAST}. These spectra were flux-calibrated with the same absolute calibration as the nebular spectra, and any potential systematic errors in the calibration are expected to cancel when calculating the (S/F$^{\star}$) ratio. The average observed flux value for the 2600 Å - 2800 Å band was $<$F$^{\star}$(2600-2800)$>$ = 3.21 x $10^{-9}$ erg cm$^{-2}$ s$^{-1}$ Å$^{-1}$, and for the 1750 Å - 1900 Å band $<$F$^{\star}$(1750-1900)$>$ = 7.37 x $10^{-9}$ erg cm$^{-2}$ s$^{-1}$ Å$^{-1}$.    

For the wavelength-integrated UV flux in the 6 eV - 13.6 eV range, we combined the SWP33895 spectrum with the $\gamma$\,Cas scattered-light spectrum obtained by the \textit{Far Ultraviolet Spectroscopic Explorer (FUSE)} to extend the spectral coverage to the Lyman limit, as discussed by \citet{2005ApJ...628..750F} (Figure \ref{FIG:6}). To obtain the far-UV flux, we binned the spectrum in 100 Å intervals and integrated it over the wavelength range corresponding to the energy range 6 eV -- 13.6 eV The observed band-integrated flux was found to be

F$^{\star}$(6.0 - 13.6) = 1.28 x $10^{-5}$ erg cm$^{-2}$ s$^{-1}$. When corrected for interstellar extinction with E(B-V) = 0.04 mag with a CCM extinction curve with $R_V$ = 3.1, F$^{\star}$(6.0 - 13.6) = 1.91 x $10^{-5}$ erg cm$^{-2}$ s$^{-1}$, and when corrected for interstellar extinction for E(B-V) = 0.08 mag,

F$^{\star}$(6.0 - 13.6) = 2.93 x $10^{-5}$ erg cm$^{-2}$ s$^{-1}$.

\subsection{Far-Infrared Intensities in IC\,59 and IC\,63}
We adopted the \textit{Herschel} PACS intensities at 70 and 160 $\mu$m and \textit{SPIRE} intensities at 250, 350, and 500 $\mu$m measured at the tips of IC\,59 and IC\,63 facing $\gamma$\,Cas, as presented by \citet{2018A&A...619A.170A}. We have reproduced their values in Table \ref{table:3}.

\begin{table}
\centering
\begin{tabular}{ |c|c|c|c|c|  }
 \hline
 \multicolumn{1}{|c|}{} &\multicolumn{2}{|c|}{IC\,63} &\multicolumn{2}{|c|}{IC\,59}  \\
 \hline
 $\lambda$($\mu$m)&$\ I_\nu$(MJy/sr)&$\delta I_\nu$&$\ I_\nu$(MJy/sr)&$\delta I_\nu$\\
 \hline
 70&611.0&6.4&33.4&2.4\\
 160&543.4&5.6&70.2&3.1\\
 250&230.7&5.6&35.2&2.9\\
 350&102.9&4.4&17.5&2.3\\
 500&40.6&4.6&7.6&1.5\\
 \hline
\end{tabular}
\caption{Far-IR photometric intensities at the tips of IC\,63 and IC\,59 adopted from \citet{2018A&A...619A.170A}.}
\label{table:3}
\end{table}

\subsection{ERE Intensities in IC\,59 and IC\,63}
\citet{2017MNRAS.469.4933L} mapped the ERE intensities across IC\,59 and IC\,63. They found maximum band-integrated ERE intensities of about 2.5 x $10^{-5}$ erg cm$^{-2}$ s$^{-1}$ sr$^{-1}$ in IC\,59 and 1.4 x $10^{-4}$ erg cm$^{-2}$ s$^{-1}$ sr$^{-1}$ in IC\,63. Spatially, the maximum ERE intensities were seen in the front regions of both nebulae facing $\gamma$\,Cas. In IC\,63, the stratification of different emission processes is more clearly discernible, in the sense that peak ERE emission occurs just behind the ionization front defined by the peak in H-$\alpha$ emission but in front of most of the pure rotational band emission from H$_{2}$ \citep{2010ApJ...725..159F} and the emissions attributed to both ionized and neutral polycyclic aromatic hydrocarbon (PAH) molecules \citep{2017MNRAS.469.4933L}. This supports an interpretation in which the ERE process traces environments dominated by photons in the 912 Å - 1100 Å wavelength range, where molecular hydrogen is either partially or totally dissociated. As long as the abundance of ERE carriers is sufficiently high relative to other absorbers of far-UV radiation, e.g H$_{2}$, to absorb the bulk of the incident photons in the 912 Å - 1100 Å range, the ERE intensity can, therefore, serve as a measure of the density of the prevailing far-UV radiation field in environments dominated by atomic hydrogen.

\section{Analysis and Results} \label{sec:analysis and results}
\subsection{Nebula - Star Colour Differences}
Denoting the nebular surface brightness as S and the corresponding stellar flux as F$^{\star}$, and defining nebular and stellar colours, respectively as  log (S$_{\lambda_1}$/S$_{\lambda_2}$) and log (F$^{\star}_{\lambda_1}$/F$^{\star}_{\lambda_2}$), we can write the colour difference \citep{1985ApJ...294..216W} as

\begin{multline}\label{equ1}
  \Delta C(\lambda_1,\lambda_2) = 
  \\
  log (S_{\lambda_1}/S_{\lambda_2}) - log (F^{\star}_{\lambda_1}/F^{\star}_{\lambda_2}) = log (S_{\lambda_1}/ F^{\star}_{\lambda_1}) - log (S_{\lambda_2}/ F^{\star}_{\lambda_2})  
\end{multline}

with $\lambda_1$ $<$ $\lambda_2$. The wavelengths $\lambda_1$ and $\lambda_2$ are the effective wavelengths of any two bands involved in the colour measurement.

\begin{figure}
    \centering
    \includegraphics[width=0.40\textwidth]{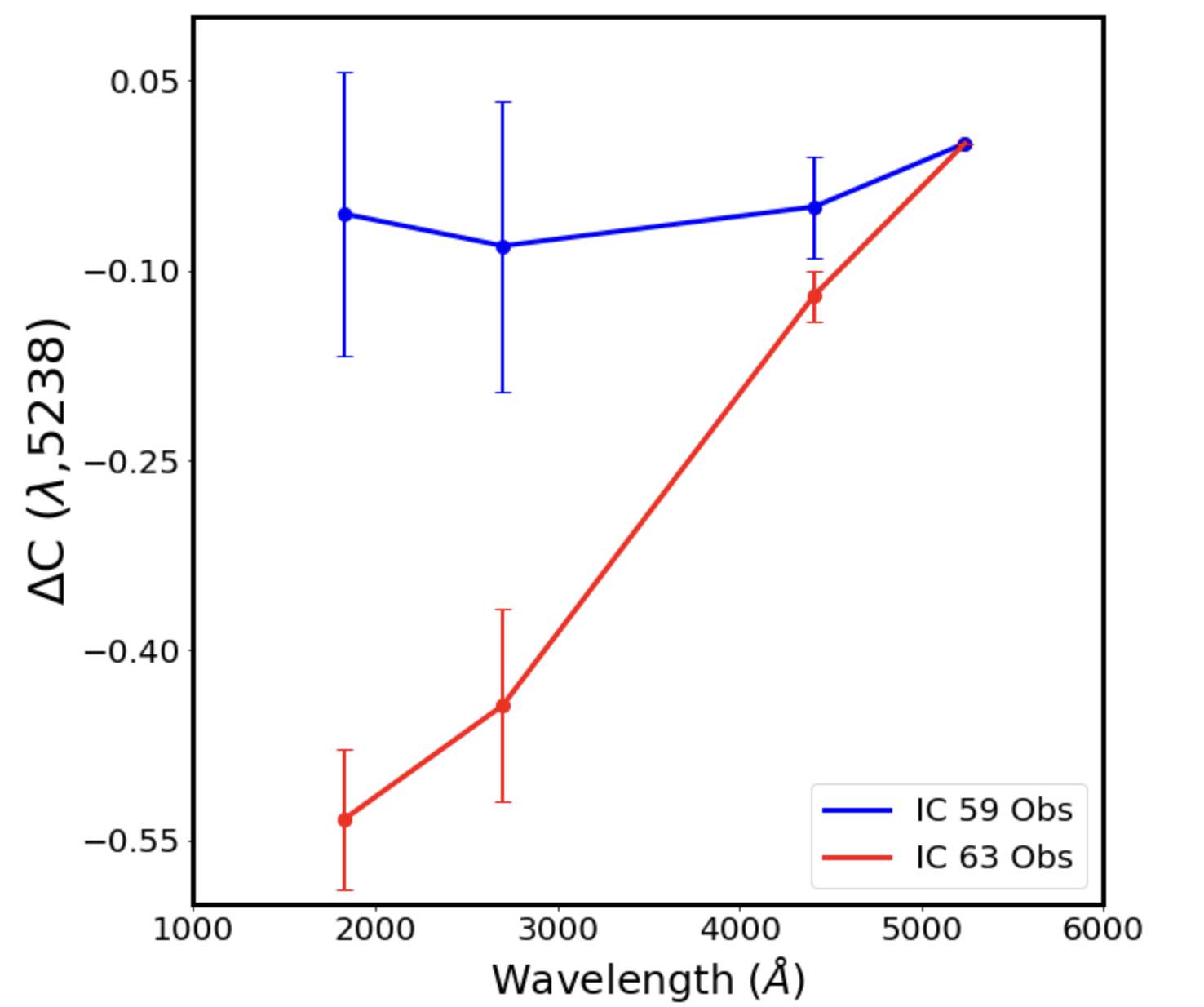}
    \caption{Nebula - Star colour differences for IC\,59 and IC\,63 relative to $\gamma$\,Cas, normalized at $\lambda_2$ = 5238 Å.}
    \label{FIG:2}
\end{figure}

The nebula - $\gamma$\,Cas colour differences for IC\,59 and IC\,63, normalized to the reference wavelength $\lambda_2$ = 5238 Å and calculated from our observed values listed in Table \ref{table:2}, are shown in Figure \ref{FIG:2}. The wavelength $\lambda_2$ = 5238 Å is the effective wavelength of the G-filter of \citet{2017MNRAS.469.4933L}. We find that IC\,63 is distinctly redder than $\gamma$\,Cas in the UV-optical wavelength range, while the colour of IC\,59 is nearly the same as that of $\gamma$\,Cas. \textit{Since there is no indication that the dust properties in the two nebulae are different from one another}, this difference is most readily explained by different effective angles under which the continuum light from $\gamma$\,Cas is being scattered toward the observer.

\citet{1985ApJ...294..216W} derived an approximate single-scattering expression for the colour difference $\Delta$C($\lambda_1$,$\lambda_2$) that includes a direct dependence on the scattering angle through the phase function of scattering. Modified for the case of an external illuminating star ($\tau_{\star}$ = 0), this expression attains the form:

% \begin{align}
%     \label{equ2}
%     \begin{split}
%         \Delta C( \lambda_1,\lambda_2) \approx  \\
%         \Phi (\alpha, g(\lambda_1))
%         % log⁡(\frac{a(\lambda_1) \times [1-exp(-\tau_0(\lambda_1))] \times exp(-\tau_1(\lambda_1)) \times \Phi (\alpha, g(\lambda_1))}{a(\lambda_2) \times [1 - exp(-\tau_0(\lambda_2))] \times exp(-\tau_1(\lambda_2)) \times \Phi (\alpha, g(\lambda_2))})
%     \end{split}
% \end{align}

\begin{multline}\label{equ2}
\Delta C( \lambda_1,\lambda_2) \approx 
\\
\log(\frac{a(\lambda_1) \times [1-\exp(-\tau_0(\lambda_1))] \times \exp(-\tau_1(\lambda_1)) \times \Phi (\alpha, g(\lambda_1))}{a(\lambda_2) \times [1 - \exp(-\tau_0(\lambda_2))] \times \exp(-\tau_1(\lambda_2)) \times \Phi (\alpha, g(\lambda_2))})
\end{multline}

where a($\lambda$) is the wavelength-dependent dust albedo, $\tau_0$($\lambda$) is the line-of-sight optical depth through the nebula at the position of the observation, $\tau_1$($\lambda$) is the optical depth encountered by the star’s light before reaching the position of observation, and $\Phi$ ($\alpha$, g($\lambda$)) is the scattering phase function for a scattering angle $\alpha$ and phase function asymmetry factor g($\lambda$). The scattering angle is defined as usual as the angle by which the scattered radiation is deflected relative to the incident beam of light. For the purpose of this study, we adopt the one-parameter \citet{1941ApJ....93...70H} (HG) phase function, with values for the dust albedo a($\lambda$) and the phase function asymmetry parameter g($\lambda$) based on empirically derived values from observations of the reflection nebulae NGC 7023 \citep{1992ApJ...395L...5W} and IC 435 \citep{1995ApJ...446L..97C}.

\begin{figure}
    \centering
    \includegraphics[width=0.40\textwidth]{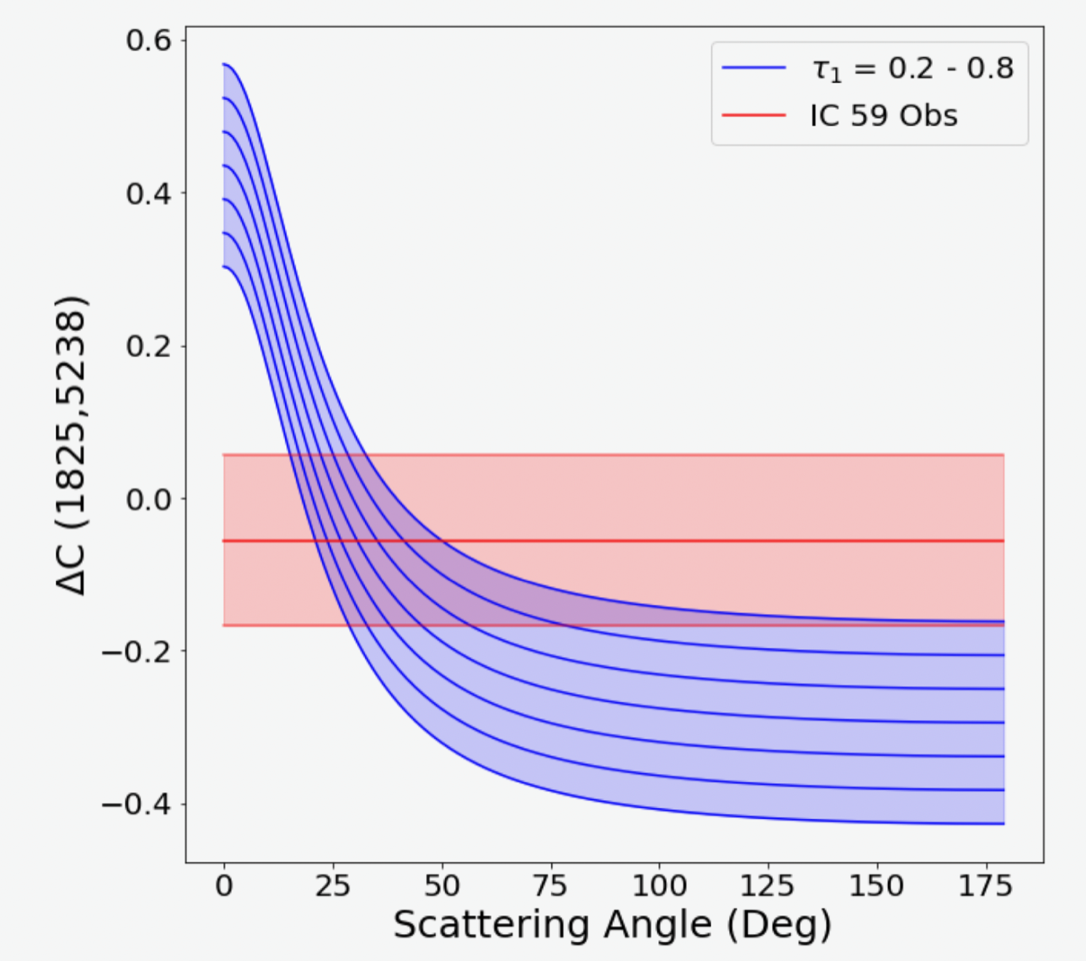}
    \caption{Predicted colour difference $\Delta$C(1825,5238), calculated with Equ. \ref{equ2} for a set of values of 0.2 $\le$ $\tau_1$(5238) $\le$ 0.8 (blue solid lines from top to bottom) is compared with the observed colour difference $\Delta$C(1825,5238) in IC\,59 (horizontal red solid line, with pink shaded area representing measurement uncertainties). The most likely scattering angle found from the intersection of the most likely colour difference (red line) with the range of predicted colour differences (blue lines)} is in the range 15$^{\circ}$ $\le$ $\alpha$ $\le$ 50$^{\circ}$, consistent with a location of IC\,59 in front of the plane of the sky containing $\gamma$\,Cas.
    \label{FIG:3}
\end{figure}

As evident from Figure \ref{FIG:2}, the largest difference in $\Delta$C($\lambda_1$,$\lambda_2$) between IC\,59 and IC\,63 occurs for $\Delta$C(1825,5238), suggesting that this value of the colour difference is most sensitive to differences in the effective scattering angles at the two nebulae. Using Equ. (\ref{equ2}), we calculated possible values of $\Delta$C(1825,5238) for IC\,59 as a function of the scattering angle, adopting a(5238) = 0.55, g(5238) = 0.60, a(1825) = 0.65, g(1825) = 0.75 \citep{1992ApJ...395L...5W, 1995ApJ...446L..97C}. We estimated $\tau_0$(5238) = 0.50, given the relatively high transparency of IC\,59 at the observed position, as judged by the number density of visible background stars. Both $\tau_0$ and $\tau_1$ were scaled to corresponding values at shorter wavelengths using a CCM extinction curve with $R_V$ = 3.7 \citep{2020ApJ...888...22V}. A comparison of predicted colour differences $\Delta$C(1825,5238) for a range 0.2 $\le$ $\tau_1$(5238) $\le$ 0.8 with the value observed in IC\,59 is shown in Figure \ref{FIG:3}. The most likely effective scattering angle is found in the range 15$^{\circ}$$\le$ $\alpha$ $\le$50$^{\circ}$, consistent with a location of IC\,59 in front of the plane of the sky containing $\gamma$\,Cas.

\begin{figure}
    \centering
    \includegraphics[width=0.40\textwidth]{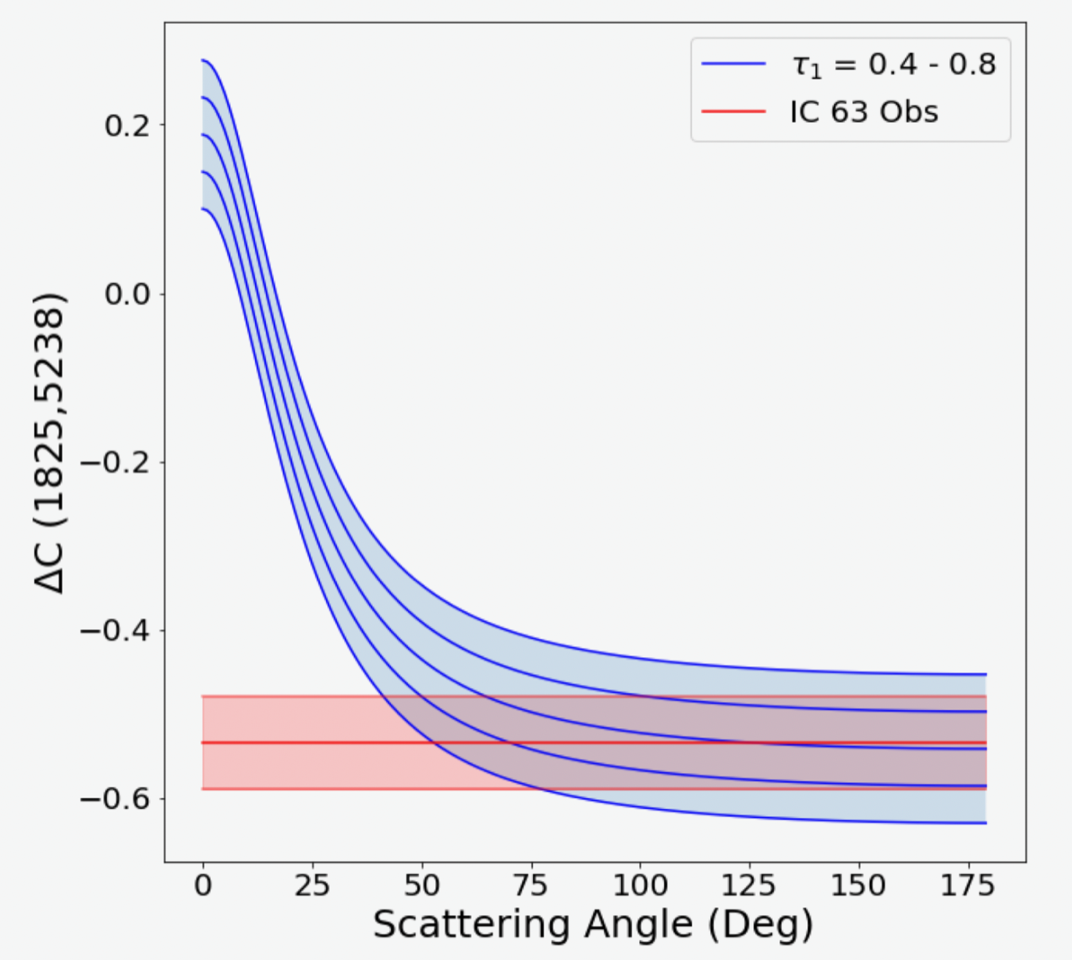}
    \caption{Predicted colour difference $\Delta$C(1825,5238), calculated with Equ. \ref{equ2} for a set of values 0.4 $\le$ $\tau_1$(5238) $\le$ 0.8 (blue solid lines from top to bottom) is compared with the observed colour difference $\Delta$C(1825,5238) in IC\,63 (horizontal red solid line, with pink shaded area representing measurement uncertainties). The most likely scattering angle is $\alpha$ > 90$^{\circ}$, as found from the intersection of the most likely colour difference (red line) with the predicted colour difference (blue lines) for the most likely value of $\tau_{1}$ = 0.6.}
    \label{FIG:4}
\end{figure}

We followed the same method to calculate the expected colour difference $\Delta$C(1825,5238) for IC\,63, but using $\tau_0$(5238) = 1.70 instead, based on extinction measurements by \citet{2020ApJ...888...22V} in the front parts of IC\,63. In Figure \ref{FIG:4}, we show a comparison of predicted colour differences as a function of the effective scattering angle $\alpha$ for a range of 0.4 $\le$ $\tau_1$(5238) $\le$ 0.8 with the observed value in IC\,63. This comparison suggests that the scattering in IC\,63 occurs most likely with $\alpha$ $>$ 90$^{\circ}$, placing IC\,63 behind the plane of the sky containing $\gamma$\,Cas, if the radial distance of IC\,63 from $\gamma$\,Cas is significantly in excess of the angular offset distance, which corresponds to about 1.0 pc.

\subsection{Dust Temperatures and Radial Distances}
\begin{figure}
    \centering
    \includegraphics[width=0.40\textwidth]{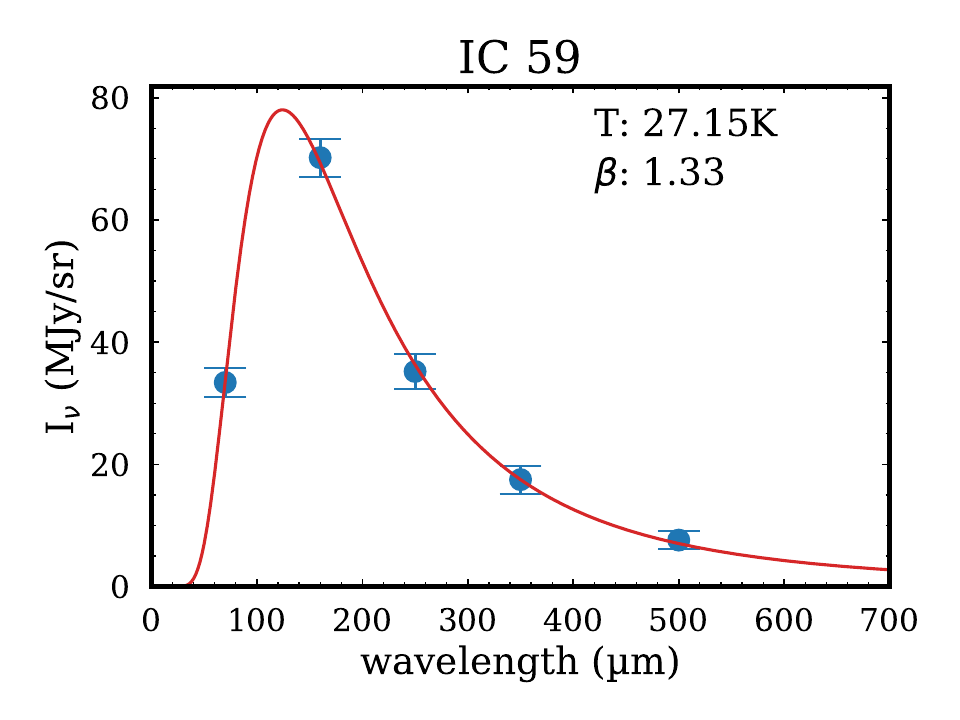}
    \caption{Modified blackbody fit to \textit{Herschel} far-IR intensities over the 70$\mu$m - 500 $\mu$m wavelength range for the tip of IC\,59 facing $\gamma$\,Cas. Data from \citet{2018A&A...619A.170A}.}
    \label{FIG:5}
\end{figure}

\begin{figure}
    \centering
    \includegraphics[width=0.40\textwidth]{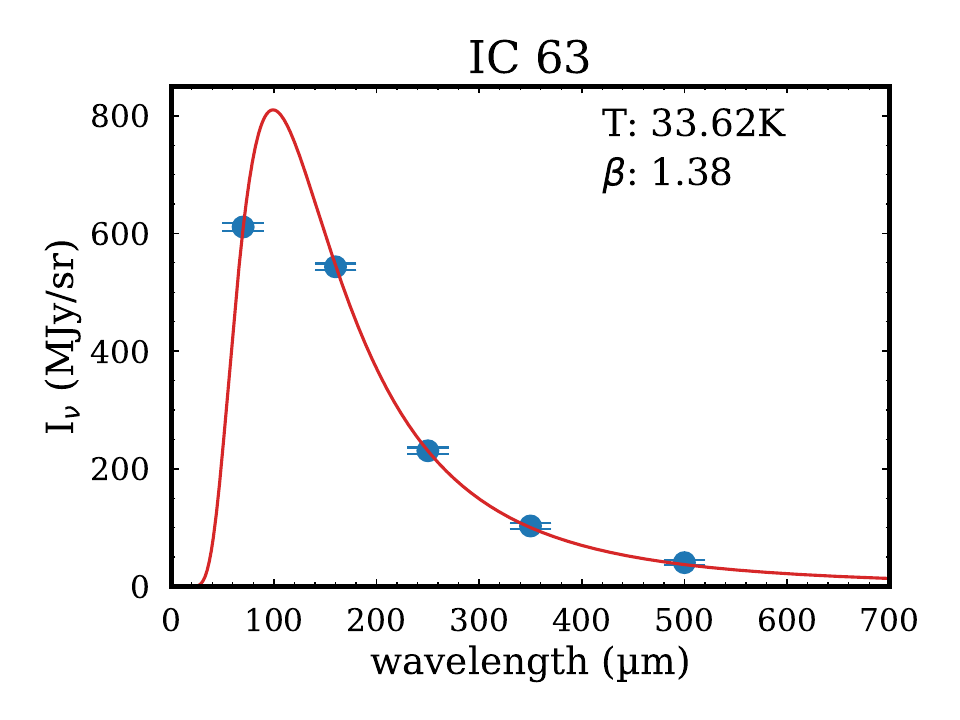}
    \caption{Modified blackbody fit to \textit{Herschel} far-IR intensities over the 70$\mu$m - 500 $\mu$m wavelength range for the tip of IC\,63 facing $\gamma$\,Cas. Data from \citet{2018A&A...619A.170A}.}
    \label{FIG:6}
\end{figure}

The equilibrium temperature of large dust grains in a radiation environment dominated by the luminosity L$_{\star}$ of a nearby star is a function of the radial distance from this star. Using the far-IR intensities (Table \ref{table:3}) observed at the tips of IC\,59 and IC\,63 by \textit{Herschel} \citep{2018A&A...619A.170A} for the wavelength range 70 $\mu$m $\le$ $\lambda$ $\le$ 500 $\mu$m, we matched the far-IR SEDs with single-temperature modified blackbody (MBB) fits, allowing both the dust temperature T$_d$ and the spectral index $\beta$ to be free parameters. The result for IC\,59 is displayed in Figure \ref{FIG:5}, that for IC\,63 in Figure \ref{FIG:6}. The best dust temperatures are T$_d$ (IC\,59) = 27.15 K and T$_d$ (IC\,63) = 33.62 K. These temperatures are well above the dust temperature found in the local diffuse interstellar medium of 20.4 $\pm$ 1.5 K \citep{2017A&A...597A.130B} or even lower \citep{2014A&A...571A..11P} and a clear indication of the dominance of the radiation from $\gamma$\,Cas in the heating of the nebular dust. The results also indicate that IC\,63 is significantly closer to $\gamma$\,Cas than IC\,59. The spectral indices $\beta$ (IC\,59) = 1.33 and $\beta$ (IC\,63) = 1.38 are consistent with a dust composition dominated by amorphous silicates and amorphous carbon \citep{2019A&A...631A..88Y} and are well within the range of values encountered in interstellar dust in the Milky Way galaxy \citep{2003A&A...404L..11D, 2017A&A...597A.130B}.

\citet{Krugel2003}, \citet{2009ApJ...698.1341U}, and \citet{Draine2011} developed detailed formalisms that relate the equilibrium temperature of large interstellar dust grains to the density of the incident radiation from a nearby star as a function of the radial distance r. The \citet{Krugel2003} formalism, which is restricted to the case of $\beta$ = 2, is not suitable for our data with $\beta$ values of 1.33 and 1.37, found respectively for IC\,59 and IC\,63. The \citet{Draine2011} formalism is developed for the more general case but with inputs involving the radiation field density rather than with explicit dependencies involving the star's luminosity and the radial distance from the star. We adopted the approach of \citet{2009ApJ...698.1341U} as the one most readily applicable to our data, as it allows for general values of the spectral index $\beta$ and provides a direct relation between the dust temperature and the star's radial distance r through the star's luminosity $L_{\star}$.

With a single star as the source of energy, the dust temperature is found as

\begin{equation}\label{equ3}
T_d(r) = \textit{K}(\beta) \times [\frac{(L_{\star} /L_{\odot})} {(r / 1000 AU)^2}]^\frac{1}{4+\beta}[\mathrm{K}]
\end{equation}

where

\begin{equation}\label{equ4}
K(\beta) = [\frac{3.89 \times 10^4} { \Gamma(4+\beta) \times \zeta (4+\beta)} \times \frac{Q(UV)} {Q(\nu_0)} \times (\frac{h\nu_0}{k})^\beta]^\frac{1}{4+\beta}
\end{equation}

Here, $\beta$ is the spectral index of the dust emissivity, $\Gamma$(4+$\beta$) and $\zeta$ (4+$\beta$))] are the Gamma and Riemann Zeta functions, Q(UV) is the stellar-flux weighted absorption efficiency of the dust grains in the UV, while $\nu_0$ = c/(125 $\mu$m) = 2.40 × $\,10^{12}$ Hz.

Considering that dust in reflection nebulae typically resides in environments with densities two or three orders of magnitude higher than those in the diffuse interstellar medium, grain growth is expected. This seems confirmed by extinction measurements by \cite{2020ApJ...888...22V} in IC 63, which revealed $R_V$ values around 3.7 in the front part of the nebula. Thus, we regarded the \citet{2003ApJ...598.1017D} interstellar dust model for $R_V$ = 4.0 to be the closest representation of the dust in IC\,59 and IC\,63, and we used \cite{2003ApJ...598.1017D} as the source for Q(UV) and Q(125 $\mu$m$)$. In conjunction with the extinction-corrected E(B-V) = 0.08 far-UV flux distribution observed from $\gamma$\,Cas, this dust model yields a ratio Q(UV)/Q($\nu_0$) = 2520.

Then, using Equ.(\ref{equ4}) with $\beta$(IC\,59) = 1.33 (Figure \ref{FIG:5}) and $\beta$(IC\,63) = 1.38 (Figure \ref{FIG:6}), we find \textit{K}(IC\,59) = 50.33 and \textit{K}(IC\,63) = 49.93.

Finally, using the best-fit dust temperatures T$_d$ (IC\,59) = 27.15 K and T$_d$ (IC\,63) = 33.62 K, together with the adopted bolometric luminosity of $\gamma$\,Cas, L$_{\star}$ = 27,150 L$_{\odot}$ \citep{2007ApJ...668..481S}, we find for the estimated radial distances from $\gamma$\,Cas: r(IC\,59) = 4.15 pc and r( IC\,63) = 2.31 pc. Combined with their offset distances of about 1.2 pc and 1.0 pc, respectively, and their locations relative to $\gamma$\,Cas, these distances imply effective scattering angles of about 17$^{\circ}$ for IC\,59 and about 154$^{\circ}$ for IC\,63, which agree well with the range of scattering angles we found in Sect. 3.1.

\subsection{Radial Distance Ratio from ERE Constraints}

Given that the ERE is excited by far-UV photons in the 10.5 eV $<$ E $<$ 13.6 eV energy range \citep{2006ApJ...636..303W, 2020Ap&SS.365...58W} and taking that the emission is isotropic, we can use measured ERE intensities to estimate the density of the prevailing far-UV radiation field in the emitting environments. The maximum ERE surface brightness in IC\,59 and IC\,63 is observed in the front parts of both nebulae facing $\gamma$\,Cas, i.e., regions exposed to the most direct and least attenuated far-UV flux from the star \citep{2017MNRAS.469.4933L}. The maximum ERE intensity in IC\,59 of I(ERE)$_{IC\,59}$ = 2.5 × $10^{-5}$ erg cm$^{-2}$ s$^{-1}$ sr$^{-1}$ persist over a significant range of corresponding scattered light intensities, covering 1.5 × $10^{-5}$ erg cm$^{-2}$ s$^{-1}$ sr$^{-1}$ $\le$ I(sca)$_{IC\,59}$ $\le$ 6.0 × $10^{-5}$ erg cm$^{-2}$ s$^{-1}$ sr$^{-1}$. This is not unexpected for an optically thin nebula that receives its illumination from behind and where higher ERE intensities generated on its back surface are partially attenuated by the nebular dust.

The character of the ERE intensity distribution in IC\,63 is very different in this respect.
The maximum ERE intensity I(ERE)$_{IC\,63}$ = 1.4 × $10^{-4}$ erg cm$^{-2}$ s$^{-1}$ sr$^{-1}$ in IC\,63 is observed in locations immediately behind the ionization front with a corresponding scattered light intensity of I(sca)$_{IC\,63}$ $\cong$ 3.5 × $10^{-4}$ erg cm$^{-2}$ s$^{-1}$ sr$^{-1}$, only to drop precipitously with increasing offset distances from $\gamma$\,Cas and decreasing scattered light intensities. 
This is consistent with illumination on the front side of the nebula as seen from Earth.

The ratio of the maximum ERE intensities I(ERE)$_{IC\,63}$ / I(ERE)$_{IC\,59}$ = 5.60 suggest a ratio of the radial distances from $\gamma$\,Cas r$_{59}$ / r$_{63}$ = (I(ERE)$_{IC\,63}$ / I(ERE)$_{IC\,59}$)$^{1/2}$ = 2.37. This is in fair agreement with the ratio r$_{59}$ / r$_{63}$ = 1.80 obtained from the dust temperatures. The distance ratio r$_{59}$ / r$_{63}$ inferred from the ratio of the ERE intensities appears larger because the observed ERE intensities in IC\,59 are most likely reduced by self-attenuation within this nebula, given that it is being illuminated by $\gamma$\,Cas from behind. By contrast, the ERE intensities in IC\,63 are observed on the directly illuminated front side and are not affected by dust attenuation. If one accepts the distance ratio r$_{59}$ / r$_{63}$ = 1.80 derived from the dust temperatures as correct, the distance ratio derived from the ratio of the ERE intensities can be made to agree with this value by invoking self-attenuation of the ERE intensities in IC\,59 by internal dust with an optical depth $\tau_{IC\,59}$ = 0.54. This optical depth is in good agreement with the values of $\tau_0$ and $\tau_1$ adopted for IC\,59 to explain the observed colour difference $\Delta$C(1825,5238) in IC\,59 (Sect. 2.1).

\subsection{Radiation Field Densities in IC\,63 and IC\,59}

Adopting the radial distances of the nebulae found from the analysis of their far-IR SEDs, r$\:_{IC\,63}$ = 2.3 pc and r$\:_{IC\,59}$ = 4.15 pc, together with the revised distance to $\gamma$\,Cas of (168 $\pm$ 4) pc and its extinction (E(B-V) = 0.08 mag) corrected far-UV flux F$^{\star}$(6.0 --- 13.6 eV) = 2.93 x $10^{-5}$ erg cm$^{-2}$ s$^{-1}$, we can estimate the corresponding far-UV flux at the surfaces of the two nebulae facing $\gamma$\,Cas. Compared to the extinction-corrected far-UV flux observed at Earth, the respective far-UV fluxes incident on the surfaces of the two nebulae are increased by the square of the ratio of the Earth -- $\gamma$\,Cas distance and the respective radial distances of the nebulae from $\gamma$\,Cas. This results in far-UV fluxes incident on IC\, 59 of 4.80  x $10^{-2}$ erg cm$^{-2}$ s$^{-1}$ and incident on IC\,63 of 1.55  x $10^{-1}$ erg cm$^{-2}$ s$^{-1}$. In Draine units of 2.7 x $10^{-3}$ erg cm$^{-2}$ s$^{-1}$, we find $G_0$(IC\,63) = 58 and $G_0$(IC\,59) = 18.

If the lower reddening of E(B-V) = 0.04 mag is accepted for $\gamma$\,Cas, even lower values are found: $G_0$(IC\,63) = 38 and $G_0$(IC\,59) = 12.

Historically, far higher UV radiation field densities had been estimated, starting with $G_0$ (IC\,63) = 650 \citep{1994A&A...282..605J}. The reduction to our current values is in part due to the actual distances of the nebulae from $\gamma$\,Cas being two to four times larger than the offset distances in the sky that had initially been taken to be the actual distances. A second factor leading to the reduction was a change in the far-UV flux from $\gamma$\,Cas seen at Earth. \citet{1994A&A...282..605J} adopted the result of \citet{1975ApJ...195..643T}, who had observed $\gamma$\,Cas during a 1971 sounding rocket flight. A flux of 1590 photons cm$^{-2}$ s$^{-1}$ Å$^{-1}$ was reported for the far-UV band 912 Å - 1075 Å, corresponding to about 3.3 x $10^{-8}$ erg cm$^{-2}$ s$^{-1}$ Å$^{-1}$. The \textit{IUE} - calibrated flux measured with \textit{FUSE} in the same spectral region is about 1.4 x $10^{-8}$ erg cm$^{-2}$ s$^{-1}$ Å$^{-1}$ \citep{2005ApJ...628..750F}, lower by about a factor 2.4. We adopted the latter for the present work, because it was obtained with a vastly more sensitive and better calibrated instrument in Earth orbit compared to the rocket experiment performed by \citet{1975ApJ...195..643T} three decades earlier.

\section{Discussion} \label{sec:Disscusion}

We employed two distinctly different methods, combining UV/optical scattering data with far-IR flux measurements, to discern the spatial location and radial distances of the nebulae IC\,59 and IC\,63 relative to their common illuminating star $\gamma$\,Cas.

The observed UV-optical colour differences between the nebulae and $\gamma$\,Cas, coupled with the wavelength dependence of the scattering phase function asymmetry, yielded a clear signal placing IC\,59 well in front of the plane of the sky containing the star, while indicating that the scattering in IC\,63 most likely occurs at scattering angles $>$ 90$^{\circ}$. Given that we relied on a single-scattering approximation, using the 1-parameter HG-phase function, we did not expect to find sufficiently precise scattering angles to infer actual star-nebula distances. This might be possible with a full multiple-scattering radiative transfer treatment with more appropriate phase functions \citep{2022A&A...659A.149B}.

Instead, we used equilibrium dust temperatures determined with modified blackbody fits to the 70 $\mu$m - 500 $\mu$m far-IR intensities observed with \textit{Herschel} to estimate the radial distances of the two nebulae from $\gamma$\,Cas, given the previously known bolometric luminosity of the star. This, coupled with the result from the colour-difference approach, places IC\,63 clearly behind the plane containing $\gamma$\,Cas, at a radial distance of about 2.3 pc.

Our result for the location of IC\,63 is in fair agreement with \citet{2013ApJ...775...84A}, who arrived, by totally different means, at the conclusion that IC\,63 lies behind $\gamma$\,Cas, with a radial distance of about 2 pc. \citet{2023ApJ...950..140C} further amplified this conclusion, while also indicating the location for IC\,59 in front of $\gamma$\,Cas, at a radial distance of 4.5 pc, in reasonable agreement with our present result based on the observed dust temperature in IC\,59.

We note that the estimate of the radial distances of the nebulae from $\gamma$\,Cas through Equ. \ref{equ3} and Equ. \ref{equ4} is dependent on a dust model through the quantity Q(UV) / (Q($\nu_0$). We adopted a \citet{2003ApJ...598.1017D} dust model with $R_V$ = 4.0 as one that most likely suits dust in comparatively dense interstellar cloud conditions found in IC\,63 and IC\,59. If we had used a \citet{2003ApJ...598.1017D} model with $R_V$ = 3.1 instead, the ratio Q(UV) / (Q($\nu_0$) would increase due to the steeper extinction in the far-UV, taking on the value Q(UV) / (Q($\nu_0$) = 2855 with the far-UV flux distribution of $\gamma$\,Cas. The parameters \textit{K}($\beta$) would increase accordingly to \textit{K}(IC\,59) = 51.10 and \textit{K}(IC\,63) = 50.72, with corresponding increases in the radial distances to r(IC\,59) = 4.23 pc and r(IC\,63) = 2.45 pc. These increases of at most about 6$\%$ over the $R_V$ = 4.0 case indicate that the results of the distance estimation do not depend in a major way on the details of the adopted dust model.

Both methods used in our analysis assume that $\gamma$\,Cas is an isotropic radiator. Not surprising for a Be star, $\gamma$\,Cas is reported to rotate at (70 $\pm$ 10) \% of its critical rotational velocity \citep{2005ApJ...624..359T}. As a result, the three-dimensional shape of $\gamma$\,Cas is expected to change from a spherical to a flattened, ellipsoidal configuration with a higher effective temperature in the polar regions compared to cooler equatorial zones. This implies that the far-UV radiative flux received at the surfaces of IC\,59 and IC\,63 may not only depend on their radial distances from $\gamma$\,Cas but also on their spatial orientation with respect to the star's rotation axis. In addition, the presence of a disk \citep{1997ApJ...479..477Q, 2012A&A...545A..59S} is likely to induce additional changes in the stellar SED when seen from different directions. Given the level of approximation involved in our two approaches, we did not attempt to include either of these complications in our analysis.

With the new constraints on the relative locations and distances of IC\,59 and IC\,63 with respect to $\gamma$\,Cas and the resulting more reliable estimates for the far-UV radiation densities prevailing in the two nebulae, we consider the $\gamma$\,Cas - nebula system an ideal laboratory for the study of PDRs in the low-UV-radiation limit. The fact that both nebulae see radiation fields of likely very similar spectral shape but different densities offers opportunities for the testing of models not found in other frequently observed PDRs. Finally, at 168 $\pm$ 4 pc, IC\,59 and IC\,63 are the nearest PDRs, offering the opportunity for superior spatial resolution with modern instruments like \textit{JWST}.

\section{Conclusions} \label{sec:conclusion}
By combining an analysis of the nebula - star colour differences over the UV-optical wavelength range with an analysis of the far-IR SEDs of the dust in IC\,59 and IC\,63, we have shown that the 3-D spatial arrangement of these two nebulae relative to their common illuminating star $\gamma$\,Cas can be resolved. We conclude:

1. IC\,63 is located behind the plane of the sky containing $\gamma$\,Cas. The tip of IC\,63 closest to the star is located at a radial distance of about 2.3 pc from the star, with scattered light from the star redirected with a scattering angle of about 154$^{\circ}$.

2. IC\,59 is in front of the plane of the sky containing $\gamma$\,Cas. The radial distance from $\gamma$\,Cas to the tip of IC\,59 closest to the star is about 4.15 pc. The effective scattering angle for IC\,59 is about 17$^{\circ}$.

3. The ratio of their radial distances is consistent with the ratio of the maximum ERE intensities observed in IC\,59 and IC\,63. This supports the utility of the ERE as a measure of the far-UV radiation density in each environment.

4. The far-UV radiation densities in the two nebulae due to irradiation by $\gamma$\,Cas, measured in \citet{1978ApJS...36..595D} units, are estimated as $G_{0}$ = 58 and $G_{0}$ = 38 for IC\,63 and $G_{0}$ = 18 and $G_{0}$ = 12 for IC\,59, with the poorly constrained interstellar reddening values of $\gamma$\,Cas of either of E(B-V) = 0.08 mag or E(B-V) = 0.04 mag being a major source of uncertainty.

5. The reduction in the far-UV radiation density in IC\,63 from $G_0$ = 650 used in earlier studies \citep{1994A&A...282..605J} to the present values is a result, in part, of the radial distance being 2.3 times larger than the offset distance in the plane of the sky, and in part due to an overestimation of the far-UV flux from $\gamma$\,Cas by \citet{1975ApJ...195..643T}.

\section*{Data Availability Statement}
The data used in this article are available from public sources, as outlined in Sect. 2.
Data generated during this work will be made available upon reasonable request to the corresponding author.

\section*{Acknowledgements}
JME gratefully acknowledges support from the University of Toledo’s Physics \textit{Research Experiences for Undergraduates} (REU) program during the summer of 2023. The University of Toledo REU program was funded through NSF Grant 1950785. ANW acknowledges fruitful interactions with Profs. Jon Bjorkman, Steven Doty, Aigen Li, and Els Peeters during various phases of this work. 
\textit{IUE} data were obtained from the \textit{Mikulski Archive for Space Telescopes at the Space Telescope Science Institute}, which is operated by the Association of Universities for Research in Astronomy, Inc.

%% Similar to \facility{}, there is the optional \software command to allow 
%% authors a place to specify which programs were used during the creation of 
%% the manuscript. Authors should list each code and include either a
%% citation or url to the code inside ()s when available.

%\software{astropy \citep{2013A$\&$A...558A..33A,2018AJ....156..123A}

%% Appendix material should be preceded with a single \appendix command.
%% There should be a \section command for each appendix. Mark appendix
%% subsections with the same markup you use in the main body of the paper.

%% Each Appendix (indicated with \section) will be lettered A, B, C, etc.
%% The equation counter will reset when it encounters the \appendix
%% command and will number appendix equations (A1), (A2), etc. The
%% Figure and Table counter will not reset.

%\appendix

%% For this sample we use BibTeX plus aasjournals.bst to generate the
%% the bibliography. The sample631.bib file was populated from ADS. To
%% get the citations to show in the compiled file do the following:
%%
%% pdflatex sample631.tex
%% bibtext sample631
%% pdflatex sample631.tex
%% pdflatex sample631.tex

\bibliographystyle{mnras}
\bibliography{mnras_template} % if your bibtex file is called example.bib

%% This command is needed to show the entire author+affiliation list when
%% the collaboration and author truncation commands are used.  It has to
%% go at the end of the manuscript.
%\allauthors

%% Include this line if you are using the \added, \replaced, \deleted
%% commands to see a summary list of all changes at the end of the article.
%\listofchanges
\bsp	% typesetting comment
\label{lastpage}
\end{document}